\documentclass[journal,twoside,web]{ieeecolor}
\usepackage{lcsys}
\usepackage{cite}
\usepackage{amsmath,amssymb,amsfonts}
\usepackage{algorithmic}
\usepackage{graphicx}
\usepackage{textcomp}

\newtheorem{rem}{Remark}
\newtheorem{thm}{Theorem}

\newtheorem{cor}{Corollary}

 \usepackage{hyperref} 
\usepackage{ulem}
\usepackage{algorithm}
\usepackage[utf8]{inputenc}
\usepackage[american]{babel}
\usepackage{csquotes}
\usepackage{mathtools}
\usepackage{siunitx}
\usepackage{tabularx}

\newcommand{\R}{\mathbb{R}}

\usepackage{pgfplots}
\usepgfplotslibrary{fillbetween}
\pgfplotsset{compat = newest}
\usetikzlibrary{arrows,positioning,shapes,intersections,patterns,calc,fit,external,decorations,decorations.markings} 
\usepackage{cite}

\def\BibTeX{{\rm B\kern-.05em{\sc i\kern-.025em b}\kern-.08em
    T\kern-.1667em\lower.7ex\hbox{E}\kern-.125emX}}
\begin{document}
\title{Distributed event-triggered flocking control of Lagrangian systems}
\author{Ernesto Aranda-Escol\'astico, Leonardo J. Colombo, Mar\'ia Guinaldo.%

\thanks{E. Aranda-Escol\'astico and M. Guinaldo have been supported by the Spanish Ministry of Science and Innovation under Projects CICYT RTI2018-094665-B-I00, RTI2018-096590-B-I00 and by Agencia Estatal de Investigación (AEI) under the project PID2020-112658RB-I00/AEI/10.13039/501100011033. L. Colombo have been founded by ``la Caixa' Foundation under the project ``Decentralized strategies for cooperative robotic swarms'' with project code LCF/BQ/PI19/11690016, and by the Spanish Ministry of Science and Innovation under Project PID2019-106715GB-C21.}

\thanks{E. Aranda-Escol\'astico is with Department of Software and Systems Engineering, Universidad Nacional de Educación a Distancia (UNED), 28040 Madrid, Spain, {\tt\small earandae@issi.uned.es}}%

\thanks{L. J. Colombo is with the Centre for Automation and Robotics (CSIC-UPM), Ctra. M300 Campo Real, Km 0,200, Arganda del Rey - 28500 Madrid, Spain, {\tt\small leonardo.colombo@car.upm-csic.es}}

\thanks{M. Guinaldo is with the Computer Science and Automatic Control Department, UNED, Juan del Rosal 16, 28040, Madrid, Spain {\tt\small mguinaldo@dia.uned.es}}%

\thanks{\textcopyright 2021 IEEE.  Personal use of this material is permitted.  Permission from IEEE must be obtained for all other uses, in any current or future media, including reprinting/republishing this material for advertising or promotional purposes, creating new collective works, for resale or redistribution to servers or lists, or reuse of any copyrighted component of this work in other works.}
}

\maketitle
\thispagestyle{empty}


\begin{abstract}
In this paper, an event-triggered control protocol is developed to investigate flocking control of Lagrangian systems, where event-triggering conditions are proposed to determine when the velocities of the agents are transmitted to their neighbours. In particular, the proposed controller is distributed, since it only depends on the available information of each agent on their own reference frame. In addition, we derive sufficient conditions to avoid Zeno behaviour. Numerical simulations are provided to show the effectiveness of the proposed control law.
\end{abstract}

\begin{IEEEkeywords}
Event-triggered control, flocking, Lagrangian systems.
\end{IEEEkeywords}

\section{Introduction}

\IEEEPARstart{F}{locking}, swarming, and schooling are common emergent collective motion behaviors exhibited in nature. These natural collective behaviors can be leveraged in multi-robot systems to safely transport large cohesive groups of robots within a workspace~\cite{rubenstein2014programmable}. To capture these effects, Reynolds introduced three heuristic rules in~\cite{reynolds1987flocks}: cohesion; alignment; and separation, to reproduce flocking motions in computer graphics. Later, these rules were used to construct flocking control algorithms. In~\cite{tanner2003stable}, the authors designed a control law that captures the following three Reynolds rules by using a collective potential function and a velocity consensus term: 
\\ \noindent$\bullet$ \textit{Cohesion}: Each agent should stay close to its neighbours.
\\ \noindent$\bullet$ \textit{Separation}: Agents cannot collide with their neighbours.
\\ \noindent$\bullet$ \textit{Alignment}: Each agent should synchronize its velocity with its neighbours.

Further, in \cite{olfati2002distributed} a theoretical framework was presented for the design and analysis of distributed flocking control algorithms based on leader-follower protocols. Since then, a great effort has been dedicated to the study of flocking control algorithms \cite{olfati2006flocking,tanner2007flocking,cucker2007emergent}.

In this work, we continue with the understanding of flocking control as satisfying the Reynolds' rules \cite{reynolds1987flocks} to study flocking control of Lagrangian systems, since they capture a large class of nonlinear control systems which appears, for instance, in robotic applications.

Despite flocking control of agents with double integrator dynamics has been widely studied in the literature \cite{anderson2}, \cite{anderson3}, \cite{sun2015rigid}, only a few authors have proposed solutions for the flocking problem of mechanical systems. In \cite{tavasoli2009flocking}, a first approach based on a gradient algorithm is proposed. Recently, flocking for uncertain Euler-Lagrange systems has been proposed in \cite{wang2013flocking,ghapani2014flocking,ghapani2016fully,li2016flocking,dong2018consensus}. The case of global connectivity maintenance was considered in \cite{mao2013distributed}; other associated problems such as input saturation \cite{yazdani2018adaptive} or actuator faults \cite{feng2019connectivity} have also been solved. However, none of those works consider the problem of communication between the agents. 

When the system under study involves a large number of agents, then the communication resources might be limited. In this context, event-triggered control has been proved to be a powerful tool to reduce the communication between the agents \cite{guinaldo2013distributed,aranda2020event,aranda2021periodic}.  Event-triggered flocking control has been previously studied in nonlinear Lipschitz systems \cite{yu2016leader,shen2019flocking,sun2019flocking}. Nevertheless, Euler-Lagrange systems provide a larger class of systems to be modeled enabling the implementation of the strategy in a wide range of areas. In this regard, there exist recent contributions handling different cooperative control objectives for Euler-Lagrange systems, such as consensus \cite{liu2016decentralised,jin2019twisting}, formation containment \cite{chen2019formation}, synchronization \cite{yao2020event} and targeted shape control \cite{aranda2021periodic}, but none of them includes stable flocking control under event-triggered communication for multi-agent Lagrangian systems. 

The main contribution of this article is the development of a novel event-triggered control protocol for stable flocking control of multi-agent Lagrangian systems, where event-triggering conditions are proposed to reduce the amount of communication required to achieve the control objective. Respect to the existing works on cooperative control of Euler-Lagrange systems, the proposed control law and trigger function neither require the knowledge of the model parameters \cite{liu2016decentralised} nor its partial linearization to implement adaptive control laws \cite{jin2019twisting,chen2019formation,yao2020event}. We further provide sufficient conditions to avoid Zeno behavior in the proposed setting with a trigger function that only depends on the local error and broadcasted states, and does not require additional variables to exclude the Zeno behavior \cite{jin2019twisting}.

We begin by reviewing the background about graph theory and forced Euler-Lagrange equations for flocking control in Section \ref{sec2}. Section \ref{sec3} addresses the problem of event-triggered control for flocking  of  Euler-Lagrange  systems and provide sufficient conditions to avoid Zeno behaviour. Finally, a numerical example with a swarm of $50$ underwater vehicles is presented in Section \ref{sec:example}.


\section{Background and problem formulation} \label{sec2}
We begin by introducing the neccessary concepts on graph theory and Lagrangian mechanics used along the paper.
\subsection{Graph theory}
Consider an \textit{undirected graph} denoted by $\mathbb{G}=(\mathcal{N}, \mathcal{E})$ where $\mathcal{N}=\{1, 2, . . . , s\}$ denotes a finite and nonempty set of nodes and $\mathcal{E}\subset\mathcal{N}\times\mathcal{N}$ a set of unordered pairs of nodes. Denote by $|\mathcal{N}|$ the cardinality of the set $\mathcal{N}$. Neighbor’s relationships of an agent $i\in \mathcal{N}$ are described  by the set $\mathcal{N}_i := \{j\in\mathcal{N} : \{j, i\}\in\mathcal{E}\}$. 

An \textit{arc} $\{j,i\}\in\mathcal{E}$ describes that nodes $i,j$ receive each other's information reciprocally. A \textit{path} between $i_1$ and $i_k$, $k\leq |\mathcal{N}|$, is a sequence of arcs of the form $\{i_1, i_2\}$, $\{i_2, i_3\},\ldots,\{i_{k-1}, i_k\}$. If each node of an undirected graph $\mathbb{G}$ has an undirected path to any other node, then $\mathbb{G}$ is said to be \textit{connected}. Besides, $A\in\mathbb{R}^{|\mathcal{N}|\times |\mathcal{N}|}$ denotes the \textit{adjacency matrix}, a matrix  $A = [a_{ij}]_{|\mathcal{N}|\times |\mathcal{N}|}$ defined by $a_{ij}>0$ if
$\{j, i\}\in\mathcal{E}$ and $a_{ij} = 0$ otherwise. Since $\mathbb{G}$ is undirected, $A$ is a symmetric matrix, i.e. $a_{ij} = a_{ji}$, for all $i, j\in\mathcal{N}$.

\subsection{Agents Dynamics: Forced Euler-Lagrange equations}\label{Sec2.3}

Consider $s\geq 2$ autonomous agents whose positions are denoted by $q_i\in\R^{d}$, and denote by $q\in\R^{d|\mathcal{N}|}$ the stacked vector of agents' positions.

The neighbor relationships between agents are described by the undirected graph $\mathbb{G}$ which is assumed to be time-invariant and connected. The stacked vector of relative positions between neighboring agents, denoted by $z\in\R^{d|\mathcal{E}|}$, is given by $z = \overline B^T q$, where $\overline B := B \otimes I_{d}\in\R^{d|\mathcal{N}|\times d|\mathcal{E}|}$, with $B$ being the incidence matrix for $\mathbb{G}$. Note that $z_k \in \R^d$ and $z_{k+|\mathcal{E}|}\in\R^d$ in $z$ correspond to $q_i - q_j$ and $q_j - q_i$ for the edge $\mathcal{E}_k$. Define $z_{k} := q_i - q_j$ and consider the desired distance between neighboring agents over the edge $\mathcal{E}_k$ as $d_{k}$.

Next, assume the motion of the agent $i\in\mathcal{N}$ is determined by a Lagrangian function $\mathbf{L}_i:\mathbb{R}^{d}\times\mathbb{R}^d\to\mathbb{R}$, that is, the Euler-Lagrange equations for $\mathbf{L}_i$ describe the dynamics for the agent. The Lagrangian function for agent $i\in\mathcal{N}$, in generalized coordiantes, is given by $\mathbf{L}_i(q_i,\dot{q}_i)=\mathbf{K}_i(q_i,\dot{q}_i)-\mathbf{U}_i(q_i)$ where $\mathbf{K}_i$ and $\mathbf{U}_i$ are the kinetic and potential energies, respectively.

While conservative forces are included into the potential energy, a non-conservative force between agents on an edge can be defined by a smooth map $\mathbf{F}_{ij}:(\mathbb{R}^{d}\times\mathbb{R}^d)\times(\mathbb{R}^d\times\mathbb{R}^{d})\to (\mathbb{R}^d\times\mathbb{R}^{d})\times(\mathbb{R}^d\times\mathbb{R}^{d})$. For instance, $\mathbf{F}_{ij}$ can describe consensus in the velocities between two agents. Lagrange-d'Alembert principle \cite{colombo2020forced} implies that the natural motions of the system are those paths $q\in\mathcal{C}^{\infty}([0,T],(\mathbb{R}^{d}\times\mathbb{R}^d))$ satisfying 
$\delta\int_{0}^{T}\mathbf{L}_i(q_i,\dot{q}_i)\,dt+\int_{0}^{T}\mathbf{F}_{ij}(q_i,q_j,\dot{q}_i,\dot{q}_j)\delta q_i\,dt=0,$ for all variations vanishing at the end points, i.e., $\delta q_i(0)=\delta q_i(T)=0$. Note that the second term is the virtual work since $\mathbf{F}_{ij}(q_i,q_j,\dot{q}_i,\dot{q}_j)\delta q_i$ is the virtual work done by the force field $\mathbf{F}_{ij}$ with a virtual displacement $\delta q_i$. Lagrange-d'Alembert principle leads to the \textit{forced Euler-Lagrange equations} \begin{equation}\label{eqforced}\frac{d}{dt}\left(\frac{\partial \mathbf{L}_i}{\partial\dot{q}_i}\right)-\frac{\partial \mathbf{L}_i}{\partial q_i}=\mathbf{F}_{ij}(q_i,q_j,\dot{q}_i,\dot{q}_j),\, i\in\mathcal{N},\,j\in\mathcal{N}_i.\end{equation}

We can expand the forced Euler-Lagrange equations, by computing the time derivative. Expanding the previous expression, equations \eqref{eqforced} takes the form
\begin{equation}\label{eq:EL}
\underbrace{\frac{\partial^2\mathbf{L}_i}{\partial\dot{q}_i\partial q_i}}_{\mathbf{C}_i(q,\dot{q})}\dot{q}_i+\underbrace{\frac{\partial^2\mathbf{L}_i}{\partial\dot{q}_{i}\partial\dot{q}_i}}_{\mathbf{M}_i(q,\dot{q})}\ddot{q}_i=\mathbf{F}_{ij}(q_i,q_j,\dot{q}_i,\dot{q}_j) +\underbrace{\frac{\partial \mathbf{L}_i}{\partial q_i}}_{\mathbf{g}_i(q)}.
\end{equation}

Equations \eqref{eqforced} determine a system of implicit second-order
differential equations. The Lagrangian $\mathbf{L}_i$ is said to be \textit{regular} (see for instance \cite{murray1994mathematical}), if for each $i\in\mathcal{N}$, the $({d|\mathcal{N}|\times d|\mathcal{N}|})$ block matrix $\mathbf{M}(q_i,\dot{q}_i)$ with blocks ${\mathbf{M}_i(q_i,\dot{q}_i)}:=\left(\frac{\partial^{2} \mathbf{L}_i}{\partial \dot q_i
\partial \dot q_i}\right)_{d\times d}$ is non-singular. In such a case, the local existence and uniqueness of solutions is guaranteed for any given initial condition. 

\begin{rem}
Note that the flocking stabilization systems (i.e., flocking control for double-integrator agents \cite{anderson2}, \cite{sun2015rigid}) can be seen as forced Euler-Lagrange equations \eqref{eqforced} by considering the Lagrangian function  $\mathbf{L}:\mathbb{R}^{d|\mathcal{N}|}\times\mathbb{R}^{d|\mathcal{N}|}\to\mathbb{R}$ given by \begin{equation}\label{stabletimedependet}\mathbf{L}(q,\dot{q})=\frac{1}{2}\sum_{i=1}^{|\mathcal{N}|}\Big(||\dot{q}
_i||^{2}-\sum_{j\in\mathcal{N}_i}V_{ij}(q_i,q_j)\Big),\end{equation} together with $\mathbf{F}_{ij}=\displaystyle{\sum_{j\in\mathcal{N}_i}(\dot{q}_j-\dot{q}_i)}$ and $\dot{q}_i=v_i$.
\end{rem}

\subsection{Problem formulation}

Consider a network given by $s\geq 2$ agents, each one with a dynamics evolving according to the Euler-Lagrange equations associated with the Lagrangian $\mathbf{L}_i:\mathbb{R}^{d}\times\mathbb{R}^{d}\to\mathbb{R}$. The network is modelled by an undirected graph $\mathbb{G}=(\mathcal{N},\mathcal{E})$ that is assumed to be connected and time-invariant.

From equations \eqref{eq:EL} we can identify the Coriolis and Mass matrices associated with the Lagrangian $\mathbf{L}_i$. These matrices must satisfy (see for instance \cite{murray1994mathematical}):

\begin{enumerate}
\item[(P1)] $\mathbf{M}_i(q,\dot{q})$ is positive definite and bounded for any $q_i\in\mathbb{R}^{d}$ and $i\in\mathcal{N}$. That is, there exists $\underline{\alpha}_i,\overline{\alpha}_i\in\mathbb{R}_{>0}$ such that $\underline{\alpha}_iI_{d\times d}\leq \mathbf{M}_i(q_i,\dot{q}_i)\leq\overline{\alpha}_iI_{d\times d}$.
\item[(P2)] $\dot{\mathbf{M}}_i(q_i,\dot{q}_i)-2\mathbf{C}_i(q_i,\dot{q}_i)$ is skew-symmetric, for each $i\in\mathcal{N}$.
\item[(P3)] $\mathbf{C}_{i}(q_i,\dot{q}_i)$ is bounded w.r.t. $q_i$ for each $i\in\mathcal{N}$ and linearly bounded w.r.t. $\dot{q}_i$. That is, there exists $\zeta_i\in\mathbb{R}_{>0}$ such that for all $i\in\mathcal{N}$, $||\mathbf{C}_i(q_i,\dot{q}_i)||\leq\zeta_i||\dot{q}_i||$.
\item[(P4)] If $\dot{q}_i$, $\ddot{q}_i \in L^\infty$, then $\frac{d}{dt}\mathbf{C}_{i}(q_i,\dot{q}_i)$ is a bounded operator.
\end{enumerate}


The goal is to show that under these conditions agents can achieve flocking motion based on Raynolds rules of alignment, cohesion and separation under an event-triggered framework. 
It is assumed that each agent is equipped with the following sensing and communication capabilities:
\begin{enumerate}
    \item [(A1)] Agent $i\in\mathcal{N}$ is able to measure the distance to its nearest neighbours continuously.
    \item[(A2)] Agent $i\in\mathcal{N}$ transmits its velocity to its neighbours at fixed instants of time, which should be determined.
\end{enumerate} 
Note that these are assumptions that agree with the reality, in the sense that many mobile robots are equipped with a set of sensors that allow them to measure its velocity and relative positions respect to other robots or obstacles. However, getting a measurement or an estimation of other agents' velocities is not easy, and flocking control requires agents to exchange this information to synchronize themselves with their nearest neighbors. Otherwise, additional assumptions over the system model and a significant increase on the computation onboard are needed \cite{chen2017adaptive,chen2019formation,yao2020event}. Though recent works have addressed the problem of achieving cooperative control objectives in multi-agent systems with position measurements only (such as in \cite{ajwad2019observer} for the consensus problem of double integrator agents),  the design of observers in cooperative Euler-Lagrange systems is generally high dimensional and complex \cite{nuno2016consensus,yang2017distributed,peng2020output}. Hence, we propose to employ an event-based strategy for the transmission of the velocity measurements in order to reduce the communication exchange.
In an event-triggered policy, the agent $j\in\mathcal{N}_i$ sends the information at instants $t_{k}^{j}$ with $k\in\mathbb{N}$ and these measurements are obtained by the agent $i\in\mathcal{N}_j$.


\section{Controller design} \label{sec3}
In this section we provide an event-triggered control law for flocking of Euler-Lagrange systems. In particular, we consider  $s\geq 2$ agents with dynamics described by the Euler-Lagrange equations as in Section \ref{Sec2.3}. 
Note that we can write equations \eqref{eq:EL} as
\begin{equation} \label{eq:EL2}
    \mathbf{M}_i(q_i)\ddot{q}_i+\mathbf{C}_i(q_i,\dot{q}_i)\dot{q}_i=\mathbf{F}_i,
\end{equation}
where $\mathbf{M}_i$ and $\mathbf{C}_i$ satisfies properties (P1)-(P4) and $\mathbf{F}_i:=\mathbf{F}_{ij}(q_i,q_j,\dot{q}_i,\dot{q}_j) +\mathbf{g}_i(q_i)$ is the control law to be designed for each agent to satisfy Raynolds rules for flocking.
According to the Reynolds model \cite{reynolds1987flocks}, the motion of each
agent in the flock is defined by the three rules of alignment,
cohesion and separation, weighted by positive constant coefficients
$\alpha_i$ and $\beta_i$. We will refer to them as Reynolds gains.

To satisfy these rules, we consider for each agent a term in the control law based on the gradient of the potential function \begin{equation}\label{Vi}V_i=\sum_{j\in\mathcal{N}_i}V_{ij}(\|z_{k}\|).\end{equation} In particular, $V_i$ must be nonnegative, of class $C^\infty$, and also its component functions $V_{ij}$ must satisfy the following properties as it was stated in \cite{tanner2003stable}:
\begin{enumerate}
    \item[(V1)] $V_{ij}\to\pm\infty$ whenever $\|z_{k}\|\to0$.
    \item[(V2)] $V_{ij}$ possesses a unique minimum and it occurs when $\|z_{k}\|$ is $d_{k}$, that is, when agents on an edge are separated at the desired distance $d_{k}$.
\end{enumerate}

Additionally, to be able to avoid Zeno behavior, we consider that $V_{ij}$ also satisfies
\begin{enumerate}
    \item[(V3)] $\nabla_{q_i}V_{ij}$ is bounded whenever $\|z_{k}\|\to\infty$.
\end{enumerate}

Under these assumptions, the acceleration of agent $i$ is determined by its neighbors
$\mathcal{N}_i$ as follows:  
\begin{equation} \label{eq:controlLawCont}
    \ddot{q}_i=-\alpha_i\sum_{j\in\mathcal{N}_i}\nabla_{q_i}V_{ij}-\beta_i\sum_{j\in\mathcal{N}_i}(\dot{q}_i(t)-\dot{q}_j(t)).
\end{equation}

\begin{rem}
Note that the first term of the control law is devoted to cohesion and separation of the agents,
while the second term is included in the control law to guarantee the alignment between agents. Note also that the choice of the gain values in the Reynolds flocking is not unique and in many situations it is application-dependent. For instance, in operations of maximal area coverage, increasing the separation gain could help to amplify the spreading of the robots. Instead, in operations that require the flock to squeeze through narrow canyons, the cohesion could be increased to make the group fit into a reduced space. 
\end{rem}

The control law \eqref{eq:controlLawCont} requires that each agent $i$ has access to the relative state $z_{k}=q_i-q_j$ and velocity $\dot{q}_j$ of its neighbors $j\in\mathcal{N}_i$. By assumption (A1), each agent $i$ can measure the distance to its neighbours, and according to (A2), it shares its velocity with them. Thus, an event-triggered mechanism is considered to reduce the communication between the agents maintaining an appropriate performance. 

If the velocity of agent $i\in\mathcal{N}$ is transmitted to its controller and its neighbours in the instants $t_k^i$, then the second factor on the left hand side of \eqref{eq:controlLawCont} becomes
\begin{equation} \label{eq:u_ve}
   -\beta_i\sum_{j\in\mathcal{N}_i}(\dot{q}_i(t_k^i)\\-\dot{q}_j(t_k^j)),
\end{equation} where $t_k^j$ is the triggering time to be defined. To determine these triggering instants, we consider the error vector
\begin{equation} \label{eq:error}
    e_i(t)=\dot{q}_i(t_k^{i})-\dot{q}_i(t),
\end{equation}
for any $t\in[t_k^{i},t_{k+1}^{i})$ such that
\begin{equation} \label{eq:tk}
    t_{k+1}^{i}=\inf\left\{ t>t_{k}:f_i(e_i(t),\dot{q}_i( t))>0\right\},
\end{equation}
where $f_i:\mathbb{R}^{d}\times\mathbb{R}^{d}\to\mathbb{R}$ is the triggering function for the $i$th agent given by
\begin{equation} \label{eq:etc}
\begin{aligned}
    f_i(t,e_i(t)) = &\|e_i(t)\|
    \\
    &-\frac{\sigma_i\sum_{j\in \mathcal{N}_{i}}\beta_i\|\dot{q}_i(t_k^i)-\dot{q}_j(t_k^j))\|^2}{2\|\sum_{j\in \mathcal{N}_{i}}\beta_i(\dot{q}_i(t_k^i)-\dot{q}_j(t_k^j))\|},
    \end{aligned}
\end{equation} where $\sigma_i\in\mathbb{R}_{>0}$ determines the number of triggered events.

Hence, the proposed control law for the coordination of multiple mechanical systems satisfying Reynold rules of flocking, based on an event-triggered protocol, is given by 
\begin{align}\label{eq:controller_flocking}
        \mathbf{F}_{ij}(q_i, q_j, \dot{q}_i, \dot{q}_j)=&-\alpha_i\sum_{j\in\mathcal{N}_i}\nabla_{q_i}V_{ij}-\frac{\partial{\mathbf{L}}}{\partial q_i}\\&-\beta_i\sum_{j\in\mathcal{N}_i}(\dot{q}_i(t_k^i)-\dot{q}_j(t_k^j)),\nonumber
\end{align} so, the control law $\mathbf{F}_i$ for agent $i\in\mathcal{N}$ is given by    
\begin{align}\label{eq:controller_flocking_dec}
 \mathbf{F}_{i}(q_i, q_j, \dot{q}_i, \dot{q}_j)=&-\alpha_i\sum_{j\in\mathcal{N}_i}\nabla_{q_i}V_{ij}\\&-\beta_i\sum_{j\in\mathcal{N}_i}(\dot{q}_i(t_k^i)-\dot{q}_j(t_k^j)).\nonumber
\end{align}

The next result shows convergence to flocking motion with the designed event triggered control law.

\begin{thm}\label{theorem}
Under the sensing and communication assumptions (A1)-(A2), a network of $s\geq 2$ agents with dynamics \eqref{eq:EL2} and control law \eqref{eq:controller_flocking_dec} achieves stable flocking if the control gain and the parameter of the event-triggering function \eqref{eq:etc} fulfill $\beta_i>0$ and $0<\sigma_i<1$ for all $i\in\mathcal{N}$, respectively. 
\end{thm}

\textit{Proof:} Consider as a candidate Lyapunov function
\begin{equation} 
    V= \sum_{i=1}^{|\mathcal{N}|}\sum_{j\in\mathcal{N}_i}\alpha_i V_{ij} +\frac{1}{2}\sum_{i=1}^{|\mathcal{N}|}\dot{q}_i^\top \mathbf{M}_i \dot{q}_i.
\end{equation}
Its time derivative is
\begin{align} 
    \dot{V}&=\sum_{i=1}^{|\mathcal{N}|}\sum_{j\in\mathcal{N}_i}\alpha_i\dot{q}_i^\top\nabla_{q_i}V_{ij}\label{eq:dV}\\&\hspace{1.7cm}+\sum_{i=1}^{|\mathcal{N}|}(\dot{q}_i^\top \mathbf{M}_i \ddot{q}_i+\frac{1}{2}\dot{q}_i^\top \dot{\mathbf{M}}_i \dot{q}_i).\nonumber
\end{align}
By using equation \eqref{eq:EL}, equation \eqref{eq:dV} can be written as
\begin{align*}
   \dot{V}&=\sum_{i=1}^{|\mathcal{N}|}\sum_{j\in\mathcal{N}_i}\alpha_i\dot{q}_i^\top\nabla_{q_i}V_{ij}+\sum_{i=1}^{|\mathcal{N}|}\frac{1}{2}\dot{q}_i^\top \dot{\mathbf{M}}_i \dot{q}_i. \nonumber \\
    &-\sum_{i=1}^{|\mathcal{N}|}\sum_{j\in\mathcal{N}_{i}}\dot{q}_i^\top[\mathbf{C}_i\dot{q}_i+\alpha_i\nabla_{q_i}V_{ij}+\beta_i(\dot{q}_i(t_k^i)-\dot{q}_j(t_k^j))]. \nonumber
\end{align*}
Using the fact that $(\dot{\mathbf{M}}_i-2\mathbf{C}_i)$ is skew-symmetric and grouping, $\alpha_i$-terms and consequently cancelling out the $\alpha_i$ terms, then $\dot{V}$ reduces to
    $\dot{V}=-\sum_{i=1}^{|\mathcal{N}|}\dot{q}_i^\top\sum_{j\in \mathcal{N}_{i}}\beta_i(\dot{q}_i(t_k^i)-\dot{q}_j(t_k^j))\big..$
Note that by using the fact that $\dot{q}_i^\top=\dot{q}_i^\top(t_k^i)-e_i^\top$, then 
\begin{equation} \label{eq:dV_bound}
    \begin{aligned}
    \dot{V}=&-\sum_{i=1}^{|\mathcal{N}|}\dot{q}_i^\top(t_k^i)\sum_{j\in \mathcal{N}_{i}}\beta_i(\dot{q}_i(t_k^i)-\dot{q}_j(t_k^j))
    \\
    &+\sum_{i=1}^{|\mathcal{N}|}e_i^\top\sum_{j\in \mathcal{N}_{i}}\beta_i(\dot{q}_i(t_k^i)-\dot{q}_j(t_k^j)).
    \end{aligned}
\end{equation}
On the one hand, since the graph is undirected and connected, the first term in \eqref{eq:dV_bound} can be rewritten such as $
    -\sum_{i=1}^{|\mathcal{N}|}\dot{q}_i^\top(t_k^i)\sum_{j\in \mathcal{N}_{i}}\beta_i(\dot{q}_i(t_k^i)-\dot{q}_j(t_k^j))=-\frac{1}{2}\sum_{i=1}^{|\mathcal{N}|}\sum_{j\in \mathcal{N}_{i}}\beta_i\|\dot{q}_i(t_k^i)-\dot{q}_j(t_k^j)\|^2.$ On the other hand, the second term can be bounded as $\sum_{i=1}^{|\mathcal{N}|}e_i^\top\sum_{j\in \mathcal{N}_{i}}\beta_i(\dot{q}_i(t_k^i)-\dot{q}_j(t_k^j))\leq \sum_{i=1}^{|\mathcal{N}|}\|e_i\|\|\sum_{j\in \mathcal{N}_{i}}\beta_i(\dot{q}_i(t_k^i)-\dot{q}_j(t_k^j)\|$.
 So, under the event-triggering condition given by \eqref{eq:etc} and from \eqref{eq:dV_bound}, we obtain that 
    \begin{align}\label{vdotbpound}
    \dot{V}\leq -\frac{1}{2}\sum_{i=1}^{|\mathcal{N}|}(1-\sigma_i)\sum_{j\in \mathcal{N}_{i}}\beta_i\|\dot{q}_i(t_k^i)-\dot{q}_j(t_k^j)\|^2\leq 0.
    \end{align}
This implies that $V$ is bounded and has a limit. Therefore, $\lim_{t\to\infty}{\|\dot{q}_i(t_k^i)-\dot{q}_j(t_k^j)\|}=0$. Observing \eqref{eq:error} and \eqref{vdotbpound}, this implies that $e_i\to0$ when $t\to\infty$ and, consequently, $\|\dot{q}_i-\dot{q}_j\|\to0$ when $t\to\infty$. Moreover, since $V$ is bounded, then $V_{ij}$ is also bounded. Thus, collisions between the interconnected agents are avoided, because of (V1), i.e, since $V_{ij}\to\pm\infty$ when $\|z_{k}\|\to0$.\hfill$\square$


In any event-triggered control framework is important to show that Zeno behaviour is avoided. Next we provide sufficient conditions to avoid Zeno behaviour. 

\begin{cor} \label{cor:cor1}
The event-triggered control law \eqref{eq:controller_flocking_dec} avoids Zeno behaviour whenever (V3) holds. 
\end{cor}

\textit{Proof:}
We next show that a minimum interevent time $0\leq t_m \leq t_{k+1}^i -t_k^i$ exists for all $i\in\mathcal{N}$. To do that, let us consider the time derivative of $e_i$ for $t\in[t_k^i,t_{k+1}^i)$,

\begin{align*}
    \frac{d}{dt}e_i(t)=-\ddot{q}_i&=\mathbf{M}_i^{-1}(q_i)\Big(\alpha_i\sum_{j\in\mathcal{N}_i}\nabla_{q_i}V_{ij}(t)\\
    &+\beta_i\sum_{j\in\mathcal{N}_i}(\dot{q}_i(t_k^i)-\dot{q}_j(t_k^j))+\mathbf{C}_i(q_i,\dot{q}_i)\dot{q}_i\Big).
\end{align*}
By properties (P1)-(P3), $\mathbf{M}_i(q_i)$ and $\mathbf{C}_i(q_i,\dot{q}_i)$ are bounded. $\dot{q}_i$ is bounded because we have shown that the Lyapunov function is bounded in Theorem \ref{theorem}. Therefore, since $\nabla_{q_i}V_{ij}(t)$ is also  bounded, then
\begin{equation}
    \frac{d}{dt}\|e_i(t)\| \leq \|\dot{e}_i(t)\| \leq l,
\end{equation}
where $l$ is a positive constant. Taking $t\in[t_k^i,t_{k+1}^i)$, we obtain $\|e_i(t)\| \leq l (t-t_{k}^i) \leq l (t_{k+1}^i-t_{k}^i)$. Thus, $t_{k+1}^i-t_{k}^i\geq \|e_i(t)\|/l$ which is larger than $0$ if $e_i(t)\neq0$. Since $e_i(t)=0$ is maintained only if the control objectives are achieved, the Zeno effect is avoided.\hfill$\square$

\section{Numerical example}\label{sec:example}

In this section, we test the distributed event-based control law \eqref{eq:controller_flocking_dec} with a network of underwater vehicles. A detailed model of the vehicles is described in \cite{aranda2021periodic}. We consider a network of $s=50$ fully actuated vehicles (rigid bodies) evolving on the special Euclidean group $SE(3)$ of rigid motions in the space. Any element of $SE(3)$ is given by $\displaystyle{g_i=\begin{bmatrix} 
R_i & b_i \\
0 & 1
\end{bmatrix}}$ with $R_i\in SO(3)$ describing the orientation for the $i^{th}$-body as a rotation matrix and $b_i=(b_i^{x},b_i^{y},b_i^{z})\in\mathbb{R}^{3}$ is the position of the center of mass for the $i^{th}$-body in inertial frame, with $\dot{b}_i=(\dot{b}_i^x,\dot{b}_i^y,\dot{b}_i^z)$ representing the velocity vector for agent $i$ in the directions $x$, $y$, $z$, respectively. The graph defining the neighbor's relations is randomly generated (ensuring that it is connected and undirected).

Besides, in some rigid body applications, the mass matrix is usually given by $\mathbf{M}_i=m_i\mathbf{I}_i$ where $m_i$ is the mass of the body and $\mathbf{I}_i$ its matrix of inertia moments. We will consider models for underwater vehicles where the elements of $\mathbf{M}_i$ may be different due to the fact that added masses have to be taken into account.

For simplicity in the model, assume that possible dissipating forces acting on the body under the water are negligible. The potential energy for the $i^{th}$ underwater vehicle is given by $U_i(R_i,b_i)=\rho\bar{\gamma}_i g\langle \bar{r}_i,R_i^{T}e_3\rangle+(\rho\bar{\gamma}_i-m_i)gb_i^{z}$, where $g$ is the gravitational acceleration, $m_i$ are the masses of each body, $\rho$, 
is the density of water, $\bar{\gamma}_i$ is the volume of each body, and $\bar{r}_i\in\mathbb{R}^3$ is a vector from the center of gravity to the center of buoyancy (in the body fixed frame) of each body. The positive $z$-axis in $\mathbb{R}^3$ for each body, i.e., $b_i^{z}$, is
taken to point downwards in the same direction as the gravity. Under these considerations, the control equations are given by
\begin{align*}
    \dot{R}_i=&R_i\hat{\Omega}_i,\quad \quad \dot{b}_i=R_i\nu_i,\\
    \mathbf{M}_i\dot{\nu}_i=&\mathbf{M}_i\nu_i\times\Omega_i-R_i^{T}(m_i-\rho\bar{\gamma}_i)ge_3+u_i\\
    \mathbf{J}_i\dot{\Omega}_i=&\mathbf{J}_i\Omega_i\times\Omega_i+\mathbf{M}_i\nu_i\times\nu_i-\rho\bar{\gamma}_i g\bar{r}_i\times(R^{T}_ie_3)+\bar{u}_i,
\end{align*}
for $i=1,2,3$, with $\Omega_i=(\Omega_i^{1},\Omega_{i}^{2},\Omega_i^{3})\in\mathbb{R}^{3}$ orientation of agent $i$ and $\hat{\Omega}$ its associated skew-symmetric matrix under the \textit{hat} isomorphism $\hat{\cdot}:\mathbb{R}^{3}\to\mathfrak{so}(3)$ \cite{murray1994mathematical}.
\begin{figure}
    \centering
    \includegraphics[width=0.75\linewidth]{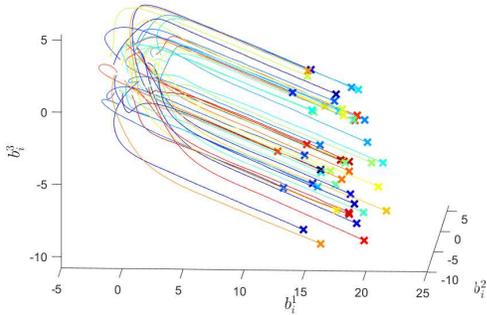}
    \caption{Trajectory in the space for the center of mass of the agents. Final states are represented by a cross ``$\times$''.}
    \label{fig:trayectory}
\end{figure}
For numerical simulations we consider all rigid bodies have mass (including added masses) $m_i = 123.8$ kg, and inertia matrices $\mathbf{M}_i=m_i\mathbf{I}_i + \hbox{diag}(65, 70, 75)$ kg, $\mathbf{J}_i=\hbox{diag}(5.46, 5.29, 5.72)$ kg$\times$m$^2$ and $\mathbf{I}_i=\hbox{Id}_{3\times 3}$ kg$\times$ m$^2$, with $\hbox{Id}_{3\times 3}$ the $(3\times 3)$-identity matrix. Also assume that $\rho\bar{\gamma}_i g=1215.8$ N and 
$\bar{r}_i = (0, 0,-0.007)^{T}$ m. We set the parameters of the control law and the event-triggering condition as $\beta_i=10$ and $\sigma_i=0.01$ for $i=1,...,50$, respectively. We choose the potential function described in \cite{cao2011distributed}, which satisfies (V1)-(V3) and whose gradient is
\begin{equation*} 
    \nabla_{q_i}V_{ij}=\begin{cases}
        (0,0,0) & \|z_{k}\|>R
        \\
        \frac{2\pi z_{k}\sin\left(2\pi\left(\|z_{k}\|-d_{k}\right)\right)}{\|z_{k}\|} & z_{k}<\|z_{k}\|\leq R
        \\
        20\frac{z_{k}}{\|z_{k}\|}\frac{\|z_{k}\|-d_{k}}{\|z_{k}\|} & \|z_{k}\|\leq d_{k}
    \end{cases},
\end{equation*}
where $d_{k}=0.5$ m and $R>\max_{k}d_{k}$ is a positive constant set to 1. Initial conditions are also randomly choosen. 
\begin{figure}
    \centering
    \includegraphics[width=0.75\linewidth]{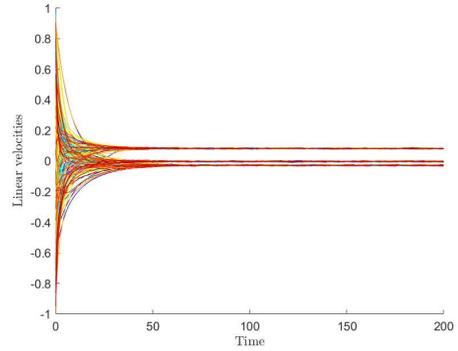}
    \caption{Linear velocities of the agents.}
    \label{fig:linear_velocity}
\end{figure}
\begin{figure}
    \centering
    \includegraphics[width=0.75\linewidth]{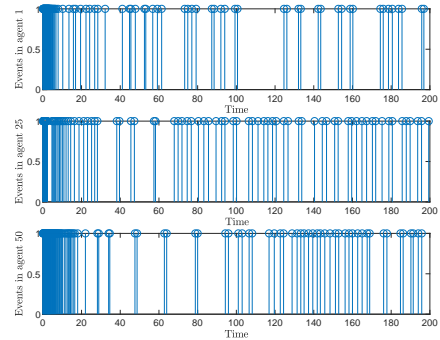}
    \caption{Example of events triggered in 3 agents.}
    \label{fig:events}
\end{figure}

In Figure \ref{fig:trayectory}, we can observe the trajectory followed by the agents. First, agents are distanced by the repulsive potential while consensus in linear velocities is achieved and the agents maintain the formation (Figure \ref{fig:linear_velocity}, the velocities of the agents converge to three values corresponding to $x,y,z$ directions). From the communication point of view, the average number of generated events is 99, i.e., each agent transmitted its velocity 99 times (in average) to their neighbours, so the average inter-event time is 2.02 s. The agent 44 generated the minimum number of transmissions (62) to its neighbours, while the agent 6 generated the maximum number (144). To illustrate the distribution of events over time, three agents have been selected (agents 1, 25 and 50), and the instances of event times are depicted as example in Figure \ref{fig:events}. At the beginning, the agents need to exchange information of their velocities very frequently (but not continuously according to Corollary \ref{cor:cor1}). However, once they are close to the consensus in velocities, the transmission of information is clearly reduced and communication resources are optimized.

\begin{figure}
    \centering
    \includegraphics[width=0.75\linewidth]{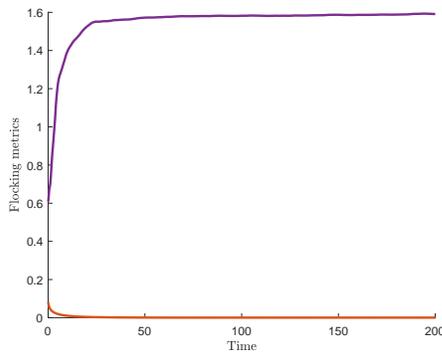}
    \caption{Flocking metrics. Average minimum distance to a neighbor in violet. Average velocity difference in orange.}
    \label{fig:metrics}
\end{figure}

We evaluate flocking behavior through the metrics described in \cite{jiahao2021learning}. The average minimum distance to a neighbor measures the cohesion between the agents. In this case, due to the nature of the potential, the distance between the agents is stabilized far enough to avoid collisions. Since the graph is fixed, we can use the average velocity difference to measure the consensus of velocities in 3D, which is clearly achieved in the example. These results are depicted in Figure \ref{fig:metrics}.

\section{Conclusions}
We have designed an event-triggered control protocol for flocking control of multi-agent Lagrangian systems. In particular, the proposed controller in this work is distributed: it only depends on the available information of each agent on their own reference frame, the velocity of its neighboors at event times, and to the desired distance to the neighboors. We have also provided sufficient condition to avoid Zeno behavior. The results show that the control objective is achieved while the amount of communication is reduced thanks to the event-triggering communication. Future work includes extension to system with delays and underactuated systems with partially unknown dynamic.

\bibliographystyle{IEEEtran}
\bibliography{mybib}

\end{document}